\documentstyle[12pt]{article}
\begin{document}
\begin{flushright}
UT-Komaba 99-1 
\end{flushright}

\begin{center} 
{\Large{\bf  Nonlocally-Correlated Disorder and 
Delocalization in One Dimension II: }}\\
\vskip 0.15cm
{\Large{\bf Localization Length }}
\vskip 1.5cm

{\Large  Ikuo Ichinose$^{\dagger}$\footnote{e-mail 
 address: ikuo@hep1.c.u-tokyo.ac.jp}and Masaomi Kimura$^{\ast}$\footnote{e-mail
 address: masaomi@ctpc1.icrr.u-tokyo.ac.jp}}  
\vskip 0.5cm
 
 $^{\dagger}$Institute of Physics, University of Tokyo, Komaba, Tokyo, 153-8902 Japan  \\
 $^{\ast}$Institute for Cosmic Ray Research, University of Tokyo, Tanashi, 
 Tokyo, 188-8502 Japan
\end{center}

\vskip 1cm
\begin{center} 
\begin{bf}
Abstract
\end{bf}
\end{center}
We study delocalization transition in a one-dimensional Dirac 
fermion system with random varying mass by using supersymmetric (SUSY) methods.
In a previous paper, we calculated density of states and found that
(quasi-)extended states near the band center are enhanced by 
nonlocal correlation of the random Dirac mass.
Numerical studies support this conclusion.
In this paper, we shall calculate localization length as a function
of correlation length of the disorder.
The result shows that the localization length is an increasing
function of the correlation of the random mass.

\newpage
\setcounter{footnote}{0}

\section{Introduction}
Random disorderd system is one of the most interesting problem
in condensed matter physics.
Especially, localization phenomenon plays an important
role in various problems, e.g., qunatum Hall effect, 
transport properties of mesoscopic systems, quantum chaos, etc.
In a previous paper (which will be refered to as paper I hereafter)\cite{IK}, 
we studied random-mass Dirac fermion in one-spatial dimension,
which is a low-energy effective field theory of random hopping
tight-binding model, quantum spin chain, etc.
This system has been studied extensively but in most of analytical studies 
only the $\delta$-function type white noise limit is considered
for technical reason\cite{DiracF,BF,CDM}.
In most of realistic systems however, disorders have nonlocal
correlations, and moreover sometimes correlation length of disorders
is larger than typical length scale of the system.
An important example is the quantum Hall state where magnetic 
length is smaller than averaged length scale of random potential.
Another example is quantum spin chain with the ground state of
dimer structure whose low-energy excitations are described by 
a Dirac fermion with random varying mass, i.e., the system 
under investigation.
There impurities generate low-energy excitations.
 
In paper I, we studied effects of nonlocal correlation of random mass,
which is quite natural for application to realistic systems,
and obtained density of states (DOS) as a function of correlation length
of the disorder.
The result indicates that (quasi-)extended states near the band
center are enhanced by nonlocal correlation of the random-mass
variables.
Recently we have performed numerical studies of the model
and verified that analytical results obtained by supersymmetric
(SUSY) methods are in good agreement with numerical calculations\cite{TTIK}.

In this paper, we shall study the same model and calculate
localization length by using the SUSY methods.
In Sect.2, we shall briefly review the model and the SUSY
methods.
In Sect.3, localization length is calculated by solving eigenvalue
problem with respect to transfer `Hamiltonian' for the spatial direction.
Section 4 is devoted for discussion.
Physical meanings of 
the result of the localization length and the DOS obtained in paper I
will be discussed.
Numerical studies also give us useful informations on the system.
As an application of the results, we shall discuss low-energy
properties of qunatum spin chains.

\setcounter{equation}{0}
\section{Model and SUSY}
In this section, we shall review the model and SUSY methods
mainly in order to fix our notations and make this paper
self-contained.
Reader who is familiar with the subjects can skip this section and 
go directly to Sect.3.
\subsection{Model and Green's functions}
We shall study random system
whose Hamiltonian is given by 
\begin{eqnarray}
{\cal H}_c &=& -\int \!\!\mbox{dx} \; \Big[\psi_R^{\dagger}i\partial_x\psi_R-
\psi_L^{\dagger}i\partial_x\psi_L  
 -im(x)(\psi_R^{\dagger}\psi_L-\psi_L^{\dagger}\psi_R)  \Big],  \nonumber   \\
 &=&-\int \!\!\mbox{dx} \; \psi^{\dagger}h \psi,   \nonumber   \\
 h &=& -i\sigma^z\partial_x+m(x)\sigma^y,  \;\; \psi=(\psi_R\psi_L)^t.
\label{Hcont}
\end{eqnarray}
 The random mass $m(x)$ is decomposed into a uniform and random piece as
\begin{equation}
m(x)=m_0+\phi(x),
\label{random} 
 \end{equation}
 where $[\phi]_{ens}=0$ and 
\begin{equation}
[\phi(x)\phi(y)]_{ens}={g \over 2\lambda}\exp (-|x-y|/\lambda),
\label{phi}
\end{equation}
with positive parameters $g$ and $\lambda$. 
The symbol $[A]_{ens}$ denotes the ensemble average of $A$ over the
randomness.
It is easily verified 
$$
\int \!\!\mbox{dx} \; [\phi(x)\phi(y)]_{ens}=g,
$$ 
and 
$$[\phi(x)\phi(y)]_{ens}
\rightarrow g\delta(x-y)\; \;  \mbox{as} \;\;\lambda \rightarrow 0.
 $$
It is obvious that the parameter $g$ controls magnitude of fluctuation
and $\lambda$ is the correlation length of the disorder. 

Single fermion Green's function at energy $\omega$ is defined as 
\begin{equation}
G_{\alpha\beta}(x,y; i\omega)=\langle x,\alpha|{1 \over h-i\omega}|y,\beta
\rangle , \;\; \alpha, \beta=R,L,
\label{Green1}
\end{equation}
where 
$|x,\alpha \rangle$ is the {\em normalized} position eigenstate 
of fermion at $x$ and 
chirality $\alpha$.
By functional integral,
\begin{eqnarray}
G_{\alpha\beta}(x,y; i\omega)&=&i\langle \psi_{\alpha}(x)
\psi^{\dagger}_{\beta}(y)
\rangle_{\psi}   \nonumber   \\
&=& {1 \over Z_{\psi}}\int {\cal D}\psi {\cal D}
\psi^{\dagger}\psi_{\alpha}(x)\psi^{\dagger}_{\beta}(y)e^{-S_{\psi}},  \nonumber \\
S_{\psi}&=& \int \!\!\mbox{dx} \; \psi^{\dagger}(ih +\omega )\psi,  \nonumber  \\
Z_{\psi}&=&\int {\cal D}\psi {\cal D}\psi^{\dagger}e^{-S_{\psi}}.
\label{Fintergal}
\end{eqnarray}
Ensemble averaged Green's function is obtained from (\ref{Green1}) as 
\begin{equation}
\bar{G}_{\alpha\beta}(x-y; i\omega)=\Big[\langle x,\alpha|{1 \over h-i\omega}|y,\beta
\rangle \Big]_{ens},
\label{Green2}
\end{equation}
where the ensemble average is taken with respect to $\phi(x)$ in (\ref{Hcont})
and (\ref{random}) according to (\ref{phi}).
Introducing the bosonic superpartner $\xi$,
\begin{equation}
\bar{G}_{\alpha\beta}(x-y; i\omega)=i\langle \psi_{\alpha}(x)\psi^{\dagger}_{\beta}(y)
\rangle_S,
\label{GSS}
\end{equation}
where
\begin{equation}
\langle {\cal A}\rangle_S=\Bigg[ \int {\cal D}\psi {\cal D}
\psi^{\dagger}{\cal D}\xi {\cal D}\xi^{\dagger} {\cal A} \;
e^{-S}\Bigg]_{ens},
\label{AS}
\end{equation}
with
\begin{equation}
S=\int\!\!\mbox{dx} \; \Big[\psi^{\dagger}(ih+\omega)\psi+
\xi^{\dagger}(ih+\omega)\xi\Big].
\label{SS}
\end{equation}

The ensemble average can be converted into the functional integral form,
\begin{eqnarray}
[\phi(x)\phi(y)...]_{ens}&=&\int {\cal D}\phi 
(\phi(x)\phi(y)...)
\exp(-S_\phi), \label{pathphi} \\
S_\phi&=&\int \!\!\mbox{dx}\,\, 
\;
\frac{1}{4g}\phi(x)(-\lambda^2{\partial_x}^2 + 1)\phi(x).
\label{Sphi}
\end{eqnarray}
From Eqs.(\ref{AS}), (\ref{SS}) and (\ref{pathphi}), the expectation value of 
operator ${\cal A}$ is given by
\begin{equation}
\langle{\cal A}\rangle_S=\int 
{\cal D}\psi{\cal D}\psi^{\dagger}{\cal D}\xi{\cal D}\xi^{\dagger}
{\cal D}\phi\mbox{ } {\cal A}\; \exp(-(S+S_\phi)).
\label{AS2}
\end{equation}
The above total action is obviously invariant under 
SUSY transformation $\psi \leftrightarrow \xi$
and has such a form that the SUSY partners $\psi$ and $\xi$ couple to
the `dynamical' real scalar field $\phi$ which is `singlet' under SUSY
transformation.

Transfer Hamiltonian is obtained by regarding
the spatial coordinate $x$ as time, and then
the system reduces to a quantum mechanical system with the two bosonic
and one fermionic variables.
By solving the Schr$\ddot{\mbox{o}}$dinger equations in the
quantum system, the Green's functions are calculated.

\subsection{Transfer Hamiltonian}
The transfer Hamiltonian is obtained by regarding the spatial
coordinate $x$ as time in the functional integral representation (\ref{AS2}).
Then the system reduces to a quantum mechanical system.
In Ref.\cite{BF}, the following canonical creation and annihilation operators
are introduced corresponding to the functional integral variables,
\begin{eqnarray}
 \psi_R  \rightarrow F_{\uparrow}, \; \; \;
\psi^{\dagger}_R \rightarrow F^{\dagger}_{\uparrow},  &&
 \psi_L \rightarrow F^{\dagger}_{\downarrow}, \;\; \;
\psi^{\dagger}_L \rightarrow F_{\downarrow}, \nonumber  \\
 \xi_R \rightarrow B_{\uparrow}, \;\;\; 
 \xi^{\dagger}_R \rightarrow B^{\dagger}_{\uparrow}, &&  
 \xi_L \rightarrow B^{\dagger}_{\downarrow}, \;\;\;
 \xi^{\dagger}_L \rightarrow B_{\downarrow}.
\label{oper}
\end{eqnarray}
Fermionic and bosonic spin operators are defined by,
\begin{eqnarray}
&& \vec{S}={1 \over 2}F^{\dagger}\vec{\sigma}F,  \nonumber \\
&& \vec{J}={1 \over 2}\bar{B}\vec{\sigma}B,
\label{spin}
\end{eqnarray}
where
\begin{equation}
\bar{B}=B^{\dagger}\sigma^z.
\label{Bbar}
\end{equation}
The commutation relations followed by $F_{\sigma}$ and $B_{\sigma}$
is 
\begin{eqnarray}
&&\{F_{\alpha}, F^{\dagger}_{\beta}\}=\delta_{\alpha\beta}, \\
&&[B_{\alpha},\bar{B}_{\beta}]=\delta_{\alpha\beta}.
\end{eqnarray}
It is proved that $\vec{S}$ and $\vec{J}$ satisfy $SU(2)$ and 
$SU(1,1)$ algebras, respectively, and they commute with SUSY charges
$Q$ and $\bar{Q}$,
\begin{equation}
Q=\bar{B}F, \; \; \bar{Q}=F^{\dagger}B,
\label{SUSYc}
\end{equation}
which satisfy 
\begin{eqnarray}\label{BF3.34}
&& Q^2=\bar{Q}^2=0, \\
&& \{Q,\bar{Q}\}=N=N_{B}+N_{F},
\end{eqnarray}
with $N_{B}=\bar{B}B$ and $N_{F}=F^{\dagger}F$.
Note that SUSY charges $Q$ and $\bar{Q}$ are different from
those in particle physics, where $\{Q,\bar{Q}\}=H$ with 
Hamiltonian $H$.
$F_{\sigma}$ and $B_{\sigma}$ 
$(\sigma=\uparrow, \downarrow)$ part of the transfer 
Hamiltonian is written
in terms of the spin operators.
Transfer Hamiltonian of $\phi$ is also obtained from Eq.(\ref{Sphi}).
The system of $\phi$ is nothing but a simple harmonic oscillator
linearly coupled with the SUSY spin ${\cal J}=S+J$.
In terms of the spin operators and the canonical boson operators of the 
harmonic oscillator $a, \; a^{\dagger}$ which correspond to $\phi$,
Hamiltonian of the system is given as,
\begin{equation} 
H=2\omega{\cal J}^{z}+2m_0{\cal J}^{x}+
\sqrt{\frac{4g}{\lambda}}{\cal J}^{x}(a+a^\dagger)+
\frac{1}{\lambda}(a^\dagger a+\frac{1}{2}).
\label{Hamiltonian0}
\end{equation}

Fermionic states of $F_{\sigma}$ are specified by 
representations of $SU(2)$.
Similarly bosonic states of $B_{\sigma}$ form multiplet of irreducible 
representations of $SU(1,1)$, which are specified by total spin 
 \begin{equation}
 J^2=(N_B^2+2N_B)/4, \;\; N_B=\bar{B}B=B^{\dagger}_{\uparrow}B_{\uparrow}
 -B^{\dagger}_{\downarrow}B_{\downarrow},
 \label{totalJ}
 \end{equation}
and $z$-component of spin $J^z$, i.e., 
\begin{eqnarray}
J^2|jn\rangle &=& j(j+1)|jn\rangle, \nonumber  \\
J^z|jn\rangle &=& \left[ {1+|2j+1| \over 2}+n\right] |jn\rangle,
\label{SU11}
\end{eqnarray}
where $j=0,\pm 1/2, \pm 1,\cdots$ and $n=0,1,\cdots$.

Quantum states of the system are specified by the quantum
numbers $N_{B}$,
$N_{F}$, $\Gamma=\bar{Q}{Q}$, $\bar{\Gamma}=Q\bar{Q}$, 
${\cal J}^z$ and $N_{\phi}=a^{\dagger}a$, which commute with
each other. The operators $\Gamma$ and $\bar{\Gamma}$ satisfy identities, 
$\Gamma^2=\bar{\Gamma}^2=0$, $\Gamma+\bar{\Gamma}=N$ and 
$\Gamma\bar{\Gamma}=\bar{\Gamma}\Gamma=0$.
The state 
$|N_{B}, N_{F}, \Gamma, \bar{\Gamma}, {\cal J}^z, N_{\phi}\rangle$
is given by the direct product 
$|N_{B}, N_{F}, \Gamma, \bar{\Gamma}, {\cal J}^z\rangle\otimes
|m\rangle_{H}$ where $N_{\phi}|m\rangle_{H}=m|m\rangle_{H}$.
In Ref.\cite{BF}, the structure of 
$|N_{B}, N_{F}, \Gamma, \bar{\Gamma}, {\cal J}^z\rangle$ 
is studied in detail. 
The state with $N_{F}=0$ is
\begin{equation}\label{BF3.52}
|2j,0,0,N,(|N+1|+1)/2+n\rangle=|jn\rangle|vac\rangle,
\end{equation}
and the one with $N_{F}=2$ 
\begin{equation}\label{BF3.53}
|2j,2,N,0,(|N-1|+1)/2+n\rangle=|jn\rangle|\downarrow\uparrow\rangle,
\end{equation}
where the fermionic sector is given as 
\begin{eqnarray}\label{BF3.38}
&&|\uparrow\rangle=F^{\dagger}_{\uparrow}|vac\rangle, \\
&&|\downarrow\rangle=F^{\dagger}_{\downarrow}|vac\rangle, \\
&&|\downarrow\uparrow\rangle=
F^{\dagger}_{\uparrow}F^{\dagger}_{\downarrow}|vac\rangle, 
\end{eqnarray}
with $|vac\rangle$, the vacuum of the fermion $F_{\sigma}$.
The states with $N_{F}=1$ can be constructed by acting 
$Q$ and $\bar{Q}$ on the above states. For $N\neq 0$, we have
\begin{eqnarray}\label{BF3.54}
&&|2j,1,(N-|N|)/2,(N+|N|)/2,|N|/2+n\rangle \nonumber \\
&&=\sqrt{\frac{n+|N|}{2n+|N|}}|jn\rangle|\downarrow\rangle
+\sqrt{\frac{n}{2n+|N|}}|jn\rangle|\uparrow\rangle, \\
&&|2j,1,(N+|N|)/2,(N-|N|)/2,|N|/2+n\rangle \nonumber \\
&&=\sqrt{\frac{n}{2n+|N|}}|jn\rangle|\downarrow\rangle
+\sqrt{\frac{n+|N|}{2n+|N|}}|jn\rangle|\uparrow\rangle.
\end{eqnarray}
For $N=0$, however, these two sets of states coincide
and are equal to
\begin{equation}\label{BF3.57}
|-1,1,0,0,n\rangle =\frac{1}{\sqrt{2}}(
|-1/2,n\rangle|\downarrow\rangle+|-1/2,n-1\rangle|\uparrow\rangle),
\end{equation}
where $n\geq 0$ and $|-1/2,-1\rangle=0$. 
Note that these are annihilated by both $Q$ and $\bar{Q}$.
In order to complete the set of the quantum space,
we introduce a set of states orthogonal to the states Eq.(\ref{BF3.57}),
\begin{equation}\label{BF3.58}
|-1,1,*,*,n\rangle = \frac{1}{\sqrt{2}}
(|-1/2,n\rangle|\downarrow\rangle-|-1/2,n-1\rangle|\uparrow\rangle),
\end{equation}
which are not eigenstates of $\Gamma$ and $\bar{\Gamma}$.

In later discussions, we shall consider `right' eigenstates of $H$,
\begin{equation}\label{BF3.59}
H|N_{B}, N_{F}, \Gamma, \bar{\Gamma},E\rangle=
E|N_{B}, N_{F}, \Gamma, \bar{\Gamma},E\rangle,
\end{equation}
which can be expanded in a basis of appropriate eigenstates of
${\cal J}^z$ and $N_{\phi}$,
\begin{equation}\label{BF3.60}
|N_{B}, N_{F}, \Gamma, \bar{\Gamma},E\rangle
=\sum_{{\cal J}^z,N_{\phi}}
\chi_{E}^{N_{B}, N_{F}, \Gamma, \bar{\Gamma}}({\cal J}^z, N_{\phi})
|N_{B}, N_{F}, \Gamma, \bar{\Gamma}, {\cal J}^z, N_{\phi}\rangle.
\end{equation}
Since $Q$ and $\bar{Q}$ do not commute with 
$N_{B}, N_{F}, \Gamma, \bar{\Gamma}$
but do with $H$, they relate degenerate eigenstates of $H$. 
For $N\neq 0$, there are two sets of SUSY doublet, 
\begin{eqnarray}\label{BF3.61}
\bar{Q}|N,0,0,N,E\rangle&=&f_{0}|N-1,1,0,N,E\rangle, \\
Q|N-1,1,0,N,E\rangle&=&\frac{N}{f_{0}}|N,0,0,N,E\rangle, \\
Q|N,0,0,N,E\rangle&=&\bar{Q}|N-1,1,0,N,E\rangle=0,
\end{eqnarray}
and 
\begin{eqnarray}\label{BF3.64}
Q|N-2,2,0,N,E\rangle&=&f'_{0}|N-1,1,0,N,E\rangle, \\
\bar{Q}|N-1,1,0,N,E\rangle&=&\frac{N}{f'_{0}}|N-2,2,0,N,E\rangle, \\
\bar{Q}|N-2,2,0,N,E\rangle&=&Q|N-1,1,0,N,E\rangle=0,
\end{eqnarray}
where  $f_{0}$ and $f'_{0}$ are constants to be determined by normalization.
For $N=0$, the space spanned by $|-1,1,*,*,n\rangle$ is not closed under the 
operation of $H$. 
The best one can do in this sector is to find 
an eigenstate of $H$ projected back onto the same sector, namely, 
\begin{equation}\label{BF3.67}
H|-1,1,*,*,E\rangle=E|-1,1,*,*,E\rangle+|\psi\rangle, 
\end{equation}
with $|\psi\rangle=\sum_{n}\psi_{n}|-1,1,0,0,n\rangle$.
Including the above state, one finds  
a superquadruplet,
\begin{eqnarray}\label{BF3.69}
&&Q|-1,1,*,*,E\rangle=f_{1}|0,0,0,0,E\rangle, \\
&&\bar{Q}|-1,1,*,*,E\rangle=f_{2}|-2,2,0,0,E\rangle, 
\end{eqnarray}  
for $N\neq0$ and 
\begin{eqnarray}\label{BF3.69-2}  
&&\bar{Q}|\mbox{ }0,0,0,0,E\rangle=f_{3}|-1,1,0,0,E\rangle, \\
&&Q|-2,2,0,0,E\rangle=f_{4}|-1,1,0,0,E\rangle, \\
&&\bar{Q}|-2,2,0,0,E\rangle=Q|0,0,0,0,E\rangle \nonumber \\
&&=\bar{Q}|-1,1,0,0,E\rangle=Q|-1,1,0,0,E\rangle=0,
\end{eqnarray}
for $N=0$.
Since the SUSY invariant `vacuum' should be annihilated
by $Q$ and $\bar{Q}$, the `vacuum' has the eigenvalues 
$N_{B}=-N_{F}=1$ and $\Gamma=\bar{\Gamma}=0$, 
and therefore, it belongs to the subspace spanned by 
$|-1,1,0,0,n\rangle\otimes|m\rangle_{H}$. 

We dedicate the rest of this section to review the `left' eigenstate of $H$, 
SUSY invariant identity operators and supertraces \cite{BF}. 
Please notice that the operator $H$ is not hermitian, 
since $J^x$ is anti-hermitian. 
`Left' eigenstate of $H$, therefore, is not identical with 
`right' eigenstate. 
Since the operator 
\begin{equation}\label{BF3.74}
U=e^{i\pi J^z+i\frac{\pi}{2}N_{F}}
\end{equation}
transforms $J^x$ as 
\begin{equation}\label{BF3.75}
U^{\dagger}J^x U=-J^x, 
\end{equation}
$U$ transforms 
\begin{eqnarray}\label{BF3.76}
U^{\dagger}HU&=&H^{\dagger}, \\
U^{\dagger}QU&=&\bar{Q}^{\dagger}, \\
U^{\dagger}\bar{Q}U&=&Q^{\dagger}. \\
\end{eqnarray}
Given a right eigenstate of $H$, $|E\rangle_{R}$, 
the state,
\begin{equation}\label{BF3.77}
|E\rangle_{L}=U^{\dagger}|E\rangle_{R},
\end{equation}
is left eigenstate of $H$, i.e., 
\begin{equation}\label{BF3.78}
\mbox{}_{L}\langle E|H=\mbox{}_{L}\langle E|E.
\end{equation}
We normalize the states as
\begin{equation}\label{BF3.79}
\mbox{}_{L}\langle E|E\rangle_{R}=1.
\end{equation}
This gives the constant $f_{0}=f'_{0}=\sqrt{N}$ for $N\neq 0$
and $f_{1}=f_{3}$, $f_{2}=f_{4}$ for $N=0$. 
For $N\neq 0$, the SUSY invariant identity operators are 
\begin{eqnarray}\label{BF3.80}
1_{N;N_{F}=0,1}=&&\sum_{E}\biggl(|N,0,0,N,E\rangle_{R}
\mbox{}_{L}\langle N,0,0,N,E|
\nonumber\\
&&+|N-1,1,N,0,E\rangle_{R}\mbox{}_{L}\langle N-1,1,N,0,E|\biggr),
\end{eqnarray}
and
\begin{eqnarray}\label{BF3.81}
1_{N;N_{F}=1,2}=&&\sum_{E}\biggl(|N-1,1,0,N,E\rangle_{R}\cdot
\mbox{}_{L}\langle N-1,1,0,N,E|
\nonumber\\
&&+|N-2,2,N,0,E\rangle_{R}\cdot\mbox{}_{L}\langle N-2,2,N,0,E|\biggr).
\end{eqnarray}
Please notice  $[Q,1]=
[\bar{Q},1]=0$.\footnote{In order to have this property,
the definition of $U=e^{i\pi{\cal J}^z+i\frac{\pi}{2}N_{F}}$ 
is essential.This is the reason why we have chosen the different
definition from that in Ref.\cite{BF}}
For $N=0$, the normalization condition is more complicated
since $|-1,1,0,0,E\rangle_{R}$ and $|-1,1,*,*,E\rangle_{R}$
are both orthogonal to their corresponding left eigenstates, 
i.e.,
\begin{eqnarray}\label{BF3.87}
&&\mbox{}_{L}\langle -1,1,0,0,E|-1,1,0,0,E\rangle_{R}=0, \\
&&\mbox{}_{L}\langle -1,1,*,*,E|-1,1,*,*,E\rangle_{R}=0. 
\end{eqnarray}
We therefore require
\begin{eqnarray}\label{BF3.89}
&&\mbox{}_{L}\langle -1,1,0,0,E|-1,1,*,*,E\rangle_{R}=1, \\
&&\mbox{}_{L}\langle -1,1,*,*,E|-1,1,0,0,E\rangle_{R}=1, 
\end{eqnarray}
with the usual normalization condition,
\begin{eqnarray}\label{BF3.85}
&&\mbox{}_{L}\langle 0,0,0,0,E|0,0,0,0,E\rangle_{R}=1, \\
&&\mbox{}_{L}\langle -2,2,0,0,E|-2,2,0,0,E\rangle_{R}=1. 
\end{eqnarray}
The corresponding identity operator for $N=0$ is therefore
\begin{eqnarray}\label{BF3.91}
1_{N=0}=&&\sum_{E}\biggl(|-2,2,0,0,E\rangle_{R}\cdot\mbox{}_{L}
\langle -2,2,0,0,E|
+|0,0,0,0,E\rangle_{R}\cdot\mbox{}_{L}\langle 0,0,0,0,E|
\nonumber\\
&&+|-1,1,*,*,E\rangle_{R}\cdot\mbox{}_{L}\langle -1,1,0,0,E|
+|-1,1,0,0,E\rangle_{R}\cdot\mbox{}_{L}\langle -1,1,*,*,E| \biggr).
\nonumber\\
\end{eqnarray}

In the following, we study the system in the long size limit, where 
the system size $=L\rightarrow\infty$.
In this case, the supertrace of an operator $Oe^{-LH}$ is expected to reduce 
to an expectation value in the SUSY invariant vacuum, 
\begin{equation}\label{BF3.93}
\langle O \rangle 
= \mbox{Str}Oe^{-LH}\rightarrow\mbox{}_{L}\langle 0|O|0\rangle_{R}e^{-LE_{b}},
\end{equation}
where $E_{b}$ is the lowest eigenvalue of $H$. 
As explained above, the vacuum $|0\rangle_R$ has eigenvalues
$N_B=-N_F=1,\Gamma=\bar{\Gamma}=0$.

\setcounter{equation}{0}
\section{Localization length}
\subsection{Eigenvalue problem}
The ensemble-averaged localization length is the parameter which measures how 
fast the avaraged 
one-particle fermion Green's function decays.
The averaged fermion Green's function is given as follows,
\begin{equation}\label{L0}
\bar{G}_{\alpha\alpha}(x;i\omega)= \, _{L}\langle 0|e^{xH}F_{\alpha}e^{-xH}
F^{\dagger}_{\alpha}|0\rangle_{R}e^{-\frac{L}{2\lambda}}.
\end{equation}
Since the state $F^{\dagger}_{\alpha}|0\rangle_{R}$ has the quantum number 
$N_B=-1$, $N_F=2$, $\Gamma=N=1$ and $\bar{\Gamma}=0$, 
\begin{eqnarray}\label{L0a}
\bar{G}_{\alpha\alpha}(x;i\omega)&=& \sum_E \; 
_{L}\langle 0|F_{\alpha}|-1,2,1,0,E \rangle_{R}
\cdot_{L}\langle -1,2,1,0,E|F^{\dagger}_{\alpha}|0 \rangle_{R}
e^{-(E-\frac{1}{2\lambda})x}e^{-\frac{L}{2\lambda}} \nonumber\\
&\simeq& e^{-\frac{x}{\xi_{\omega}}}, 
\end{eqnarray}
where $\xi_{\omega}$ is `imaginary-time' localization 
length and the state $|-1,2,1,0,E\rangle_{R}$ satisfies the 
following equation,
\begin{equation}\label{L1}
H|-1,2,1,0,E\rangle_{R}=E|-1,2,1,0,E\rangle_{R}.
\end{equation}
It is obvious that the localization length is related with the 
eigenvalue $E$.
We shall obtain the ginuine localization length 
$\xi_{\epsilon}$ from $\xi_{\omega}$ 
by the analytic continuation $i\omega \rightarrow \epsilon$.
In the following, we rescale the paremeters as $\omega\rightarrow
g\omega$, $m_0\rightarrow g m_0$ and
$\lambda\rightarrow \frac{\lambda}{g}$. 

$|-1,2,1,0,E\rangle_{R}$ is given by a linear 
conbination of the states 
$|n\rangle_{1}\equiv |-1,2,1,0,n\rangle_{R}$, and can be written as 
\begin{equation}\label{L2}
|-1,2,1,0,E\rangle_{R}
=\sum_{n=0}^{\infty}\chi_{n,m}|n\rangle_{1}|m\rangle_{H},
\end{equation}
with coefficients $\chi_{n,m}$.
Reminding that the wave function $\{\phi_{n,m}\}$ for $|0\rangle_R$,
$$
|0\rangle_R=\sum_{n,m}\phi_{n,m}|-1,1,0,0,n\rangle\otimes|m\rangle_H,
$$
has the form\cite{IK},
\[\phi_{n,m}=
\lambda^{\frac{m}{2}}(\phi^{[1]}_{n,m}+\lambda\phi^{[2]}_{n,m}),\]
where $\phi^{[i]}_{n,m}(i=1,2)$ are independent of $\lambda$,
and the `energy eigenvalue' of $|0\rangle_R$ is $\frac{1}{2\lambda}$,
we assume that $\chi_{n,m}$ and $E$ are expanded as
\begin{eqnarray}\label{L3}
\chi_{n,m}&=&\lambda^{\frac{m}{2}}
(\chi^{[0]}_{n,m}+\lambda\chi^{[1]}_{n,m}+\lambda^2\chi^{[2]}_{n,m}\cdots),\\
E&=&\frac{1}{2\lambda}+E^{[0]}+\lambda E^{[1]}+\lambda^2 E^{[2]}\cdots,
\end{eqnarray}
where $\chi^{[i]}_{n,m}$ and $E^{[i]}$ are independent of $\lambda$.
Equation (\ref{L1}) gives the Shr$\ddot{\mbox{o}}$dinger equation of $\chi_n$;
\begin{eqnarray}\label{L4}
&&\sum_{i,n}\Bigl(H_{0}+4({\cal J}^x)^2+\frac{m}{\lambda}\Bigr)
\lambda^{i+\frac{m}{2}}\chi^{[i]}_{n,m}|n\rangle_{1} \nonumber \\
&&+\sqrt{\frac{4}{\lambda}}{\cal J}^x\sum_{i,n}
(\sqrt{m}\lambda^{i+\frac{m-1}{2}}\chi^{[i]}_{n,m-1}+
\sqrt{m+1}\lambda^{i+\frac{m+1}{2}}\chi^{[i]}_{n,m+1})|n\rangle_{1}
\nonumber \\
&&=\sum_{j}\lambda^{j}E^{[j]}
\sum_{i,n}\lambda^{i+\frac{m}{2}}\chi^{[i]}_{n,m}|n\rangle_{1},
\end{eqnarray}
where 
$$
H_0=2\omega{\cal J}^z+2m_0{\cal J}^x-2({\cal J}^x)^2.
$$
Comparing terms of each order of $\lambda$ leads us to
the following series of equations,  \\
the $O(\lambda^{-1})$ terms;
\begin{equation}\label{L5}
\sum_{n}\chi^{[0]}_{n,m}|n\rangle_{1}+
\sqrt{\frac{4}{m}}{\cal J}^{x}\sum_{n}\chi^{[0]}_{n,m-1}|n\rangle_{1}
=0,
\end{equation}
the $O(\lambda^{0})$ terms;
\begin{eqnarray}\label{L6}
&&\sum_{n}H_{0}\chi^{[0]}_{n,m}|n\rangle_{1} \nonumber\\
&&+4({\cal J}^{x})^2\sum_{n}\chi^{[0]}_{n,m}|n\rangle_{1}
+2\sqrt{m+1}{\cal J}^{x}\sum_{n}\chi^{[0]}_{n,m+1}|n\rangle_{1}\nonumber\\
&&+m\sum_{n}\chi^{[1]}_{n,m}|n\rangle_{1}
+2\sqrt{m}\sum_{n}\chi^{[1]}_{n,m}|n\rangle_{1}\nonumber\\
&&=E^{[0]}\sum_{n}\chi^{[0]}_{n,m}|n\rangle_{1},
\end{eqnarray}
the $O(\lambda^{1})$ terms;
\begin{eqnarray}\label{L7}
&&\sum_{n}H_{0}\chi^{[1]}_{n,m}|n\rangle_{1} \nonumber\\
&&+4({\cal J}^{x})^2\sum_{n}\chi^{[1]}_{n,m}|n\rangle_{1}
+2\sqrt{m+1}{\cal J}^{x}\sum_{n}\chi^{[1]}_{n,m+1}|n\rangle_{1}\nonumber\\
&&+m\sum_{n}\chi^{[2]}_{n,m}|n\rangle_{1}
+2\sqrt{m}\sum_{n}\chi^{[2]}_{n,m}|n\rangle_{1}\nonumber\\
&&=E^{[0]}\sum_{n}\chi^{[1]}_{n,m}|n\rangle_{1}
+E^{[1]}\sum_{n}\chi^{[0]}_{n,m}|n\rangle_{1},
\end{eqnarray}
and so on.
It is easily seen that
Eq.(\ref{L5}) and Eq.(\ref{L6}) for $m=0$ determine $\chi^{[0]}_{n,m}$. 
Equation(\ref{L6}) for $m=0$ is identical with the equation of $\chi_{n}$
which is considered by Balents and Fisher \cite{BF}.
In a similar way, Eq.(\ref{L6}) and Eq.(\ref{L7}) for $m=0$ determine
$\chi^{[1]}_{n,m}$ and $E^{[1]}$.
\subsection{Solution in O($\lambda^0$)}

The solution $\chi^{[0]}_{n,m}$ and $E^{[0]}$ are given by Balents and Fisher
\cite{BF}. 
We review the derivation of them again for completeness.
Equation to be solved is 
\begin{equation}\label{L8}
(2\omega{\cal J}^z-4M\omega{\cal J}^x+4(\omega{\cal J}^x)^2))
\sum_{n}\chi^{[0]}_{n,0}|n\rangle_1=E^{[0]}\sum_{n}\chi^{[0]}_{n,0}|n\rangle_1,
\end{equation}
with $M=2m_0$. 
Using the representation of ${\cal J}^z$ and 
${\cal J}^x$ for $|n\rangle_1$,
\begin{eqnarray}\label{L9}
{\cal J}^z|n\rangle_1=(n+\frac{1}{2})|n\rangle_1 \\
{\cal J}^x|n\rangle_1=\frac{n+1}{2}|n+1\rangle_1 -\frac{n}{2}|n-1\rangle_1,\\
\end{eqnarray}
we have the descrete equation of $\chi^{[0]}_{n,0}$,
\begin{eqnarray}\label{L10}
&& 2\omega(n+\frac{1}{2})\chi^{[0]}_{n,0}+2M(n\chi^{[0]}_{n-1,0}-
(n+1)\chi^{[0]}_{n+1,0})
-(n+2)(n+1)\chi^{[0]}_{n+2,0}   \nonumber \\
&& +(2n^2+2n+1)\chi^{[0]}_{n,0} -n(n-1)\chi^{[0]}_{n-2,0}
=E^{[0]}\chi^{[0]}_{n,0}.
\end{eqnarray}
Since multiplying a solution for $M$ by $(-1)^n$ yields a 
solution for $-M$, we expect 
\begin{equation}\label{L10-1}
\chi^{[0]}_{n,0}=c_3\chi^{[0]}(n,M,E)+c_4(-1)^n \chi^{[0]}(n,-M,E),
\end{equation}
where $\chi(n,M,E)$ is obtained by solving the continuum equation 
derived from Eq.(\ref{L10}),
\begin{equation}\label{L11}
\biggl(2\omega n-2M\bigl(2n\frac{\partial}{\partial n}+1\bigr)
-(2n\frac{\partial}{\partial n}+1\bigr)^2\biggr)\chi^{[0]}(n,M,E)
=E^{[0]}\chi^{[0]}(n,M,E).
\end{equation}
Changing variables $n$ to $z=\ln n$, Eq.(\ref{L11}) gives
\begin{equation}\label{L12}
\Bigl(-(2\frac{\partial}{\partial z}+1)^2-2M(2\frac{\partial}{\partial z}+1)
+2\omega e^z
\Bigr)\Phi_E(z,M)=E^{[0]}\Phi_E(z,M),
\end{equation}
where $\Phi_E(z,M)=\chi^{[0]}(n,M,E)$.

Though it is possible to solve Eq.(\ref{L12}) exactly, 
we use the hard-wall approximation for simplicity, 
i.e., the `potential' term $\omega e^z$ 
is neglected and the boundary condition $\Phi_E(|\ln\omega|,M)=0$ is 
imposed instead.
Then the solution is obtained as
\begin{equation}\label{L13}
\Phi_E(z,M)=e^{-\frac{1+M}{2}z}\sin\frac{\beta(z-z_\omega)}{2},
\end{equation}
with $\beta=\sqrt{E-M^2}$ and $z_\omega=|\ln\omega|$.
In order to obtain a solution for $\chi^{[0]}_{n,0}$, namely, 
to fix the constants $c_3$ and $c_4$ in Eq.(\ref{L10-1}), 
additional conditions have to be imposed. 
Since neither $\chi^{[0]}(n,M,E)=1$ 
nor $(-1)^n$ are, unfortunately, solutions in the limit 
$\omega=M=0$, we are
not able to use similar boundary conditions used in paper I 
in which we obtained the DOS. 

Our strategy is to compare the solution 
(\ref{L13}) with the one obtained by solving 
the discrete equation (\ref{L10}) by neglecting the term 
$\omega{\cal J}^z$.
Since the other terms in the Hamiltonian except $\omega{\cal J}^z$ 
depend on only 
${\cal J}^x$, we consider eigenstate $|\alpha\rangle$ of ${\cal J}^x$;
\begin{equation}\label{L14}
{\cal J}^x |\alpha\rangle = \alpha|\alpha\rangle.
\end{equation}
This state is the eigenstate of $H$ with $\omega=0$, because 
$$
 H|\alpha\rangle=(4M\alpha-4\alpha^2)|\alpha\rangle=E|\alpha\rangle. 
$$
Expanding $|\alpha\rangle$ in terms of $|n\rangle_1$,
\begin{equation}\label{L15}
|\alpha\rangle = \sum_n\psi_n(\alpha)|n\rangle_1,
\end{equation}
we find $\chi^{[0]}_{n,0}=c_{+}\psi_n(\alpha_{+})+c_{-}\psi_n(\alpha_{-})$
with $\alpha_{\pm}=\frac{M\pm i\beta}{2}$. This approximation is valid for 
$n \ll 1/\omega$.
Eq.(\ref{L15}) leads us to the following equation of $\psi_n(\alpha)$,
\begin{equation}\label{L16}
(n+1)\psi_{n+1}(\alpha)-n\psi_{n-1}(\alpha)=-2\alpha\psi_{n}(\alpha).
\end{equation}
Equation (\ref{L16}) can be solved easily by introducing the function 
$\hat{\psi}(w,\alpha)\equiv \sum_n \psi_n(\alpha)w^n$. 
By using the identity for $\sigma(w)=\sum_n \sigma_nw^n$,
\begin{eqnarray}\label{L16-1}
\frac{1}{2}\sum_n ((n+1)\sigma_{n+1}-n\sigma_{n-1})w^n
&=&\frac{1}{2}((1-w^2)\frac{\partial}{\partial w}-w)\sigma(w) \nonumber \\
&\equiv& D_{w}\sigma(w),
\end{eqnarray} 
the difference equation (\ref{L16}) is converted into 
the differential equation,
$$
\Big((1-w^2)\frac{\partial}{\partial w}-w+2\alpha\Big)\hat{\psi}(w,\alpha)=0,
$$ 
and solution to this equation is obtained as
$$
\hat{\psi}(w,\alpha)=c(1-w)^{-\frac{1}{2}+\alpha}(1+w)^{-\frac{1}{2}-\alpha},
$$ 
with $c$ being a constant. 
Thus we have 
\begin{eqnarray}\label{L17}
\psi_n(\alpha)&=&
\frac{c}{n!}\oint_C \frac{dw}{2\pi i}
\frac{(1-w)^{-\frac{1}{2}+\alpha}(1+w)^{-\frac{1}{2}-\alpha}}{w^{n+1}}
\nonumber\\
&=&\frac{c}{\Gamma(n+1)}\frac{\cos\pi\alpha}{\pi}\sum_{l=0}^{n}(-1)^{l}
\Gamma(\frac{1}{2}-l+n-\alpha)\Gamma(\frac{1}{2}+l+\alpha).
\end{eqnarray}
The contour $C$ is the circle whose center is located at
the origin and radius is small enough.
Dominant terms in Eq.(\ref{L17}) are $l=0$ and/or $l=n$ for large $n$.
Using the fomula of the Gamma function,
\begin{equation}\label{L18}
\lim_{n\rightarrow\infty}\frac{\Gamma(n+a)}{\Gamma(n)n^a}=1, 
\end{equation}
we find that, for large $n$, Eq.(\ref{L17}) can be approximated as, 
\begin{equation}\label{L19}
\psi_n(\alpha)=n^{-\frac{1}{2}-\alpha}\frac{\Gamma(\frac{1}{2}+\alpha)}
{\sqrt{\pi}}
+(-1)^n n^{-\frac{1}{2}+\alpha}\frac{\Gamma(\frac{1}{2}-\alpha)}{\sqrt{\pi}},
\end{equation}
where we have fixed $c=\frac{\sqrt{\pi}}{\cos\pi\alpha}$ without 
loss of generality. 
As we shall see, in the limit $M\rightarrow 0$ and 
$\omega\rightarrow 0$, $\alpha_{\pm}$ tend to vanish, and thus 
$\Gamma(\frac{1}{2}\pm\alpha)$ tend to $\sqrt{\pi}$.
Therefore we have
\begin{equation}\label{L20}
\chi^{[0]}_{n,0}=
e^{-\frac{1+M}{2}z}(c_{+}e^{-i\frac{\beta}{2}z}+c_{-}e^{i\frac{\beta}{2}z})
+(-1)^n 
e^{-\frac{1-M}{2}z}(c_{+}e^{i\frac{\beta}{2}z}+c_{-}e^{-i\frac{\beta}{2}z}).
\end{equation}

We shall compare Eq.(\ref{L20}) with Eq.(\ref{L10-1}).
These two solutions should coincide with 
each other for $1/\omega \ll n\ll 1$.
This matching condition gives
\begin{eqnarray}
 c_{+}=c_{-}=c_{3}/2=c_{4}/2, && \;  \beta z_\omega/\pi=\mbox{an even integer},
 \nonumber  \\
 c_{+}=-c_{-}=c_{3}/2i=-c_{4}/2i, && \; 
 \beta z_\omega/\pi=\mbox{an odd integer}
\end{eqnarray} 
From the above condition on $\beta z_\omega/\pi$, $E^{[0]}$ is determined. 
So much is the review for the zeroth order solutions.

We summarize the solutions $\chi^{[0]}$ and the eigenvalues $E^{[0]}$ of 
Eq.(\ref{L8}) as follows;
\begin{eqnarray}\label{L21}
\chi^{[0]}_{n,0}&=&
c_{3}\Bigl(e^{-\frac{1+M}{2}z}\sin\frac{k\pi}{2z_\omega}(z-z_\omega)
+(-1)^{n+k} e^{-\frac{1-M}{2}z}
\sin\frac{k\pi}{2z_\omega}(z-z_\omega)\Bigr), \\
E^{[0]}&=& M^2+\Bigl(\frac{\pi k}{z_\omega}\Bigr)^2, 
\end{eqnarray}
with an integer $k>1$. The constant $c_3$ is determined 
to be $z_\omega^{-1/2}$ 
by the normalization condition. 
By solving Eq.(\ref{L5}),
$\chi^{[0]}_{n,m}(m>0)$ is determined in terms of $\chi^{[0]}_{n,0}$.
\subsection{Solution in O($\lambda$)}
Now let us turn to $E^{[1]}$, which gives the first-order correction
of the localization length.  
From 
Eq.(\ref{L6}) and Eq.(\ref{L7}) for $m=0$, we have,
\begin{equation}\label{L22}
(H_0-E^{[0]})\sum_{n}\chi^{[1]}_{n,m}|n\rangle_{1}
=-\frac{4}{m+1}{\cal J}^x(H_0-E^{[0]}){\cal J}^x
\sum_{n}\chi^{[0]}_{n,m}|n\rangle_{1}
+E^{[1]}\sum_{n}\chi^{[0]}_{n,m}|n\rangle_{1}.
\end{equation}
Since $E^{[1]}$ does not depend on $m$, it is sufficient to solve
Eq.(\ref{L22}) for $m=0$ in order to obtain $E^{[1]}$. 
We employ a similar strategy used above, namely, we shall obtain 
solutions by both the hard-wall approximation to the 
continuum equation and 
the approximation which neglects the term $\omega {\cal J}^z$
in $H_0$ in the original descrete equation.
The solutions depend on the constant $E^{[1]}$. 
We obtain $E^{[1]}$ by matching the solutions obtained by these 
methods in the region where the above two approximations are both legitimate.
Using the representaion of ${\cal J}^x$ and ${\cal J}^z$, Eqs.(\ref{L9})
and (\ref{L22}) give 
\begin{eqnarray}\label{L23}
&& \Bigl(-(2\frac{\partial}{\partial z}+1)^2
-2M(2\frac{\partial}{\partial z}+1)
+2\omega e^z-E^{[0]}
\Bigr)\Psi_E(z,M) \nonumber \\
&& \; \; =-4\omega e^z(2\frac{\partial}{\partial z}+1)\Phi_E(z,M)
+E^{[1]}\Phi_E(z,M),
\end{eqnarray}
in the continuum approximation with $n=e^z$ and 
$$
\Psi_E(z,M)=\chi^{[1]}(e^z, M)+4(2\frac{\partial}{\partial z}+1)^2\Phi_E(z,M),
$$
where 
$$
\chi^{[1]}_{n,0}=c_3\chi^{[1]}(n, M)+c_4(-1)^n \chi^{[1]}(n,-M).
$$
In the hard-wall approximation, we neglect the term $\omega e^z$ on the left 
hand side of Eq.(\ref{L23}), and we impose the boundary condition 
$\Psi_E(z_{\omega},M)=0$ instead. 
Making use of Eq.(\ref{L13}) and assuming  
$\Psi_E(z,M)=e^{-\frac{1+M}{2}z}\eta(z,M)$,
\begin{equation}\label{L24}
(-4\frac{\partial^2}{\partial z^2}-\beta^2)\eta(z,M)
=-8\omega e^z(\frac{\partial}{\partial z}-\frac{M}{2})
\sin\frac{\beta}{2}(z-z_{\omega})+E^{[1]}\sin\frac{\beta}{2}(z-z_{\omega}).
\end{equation}
It is not so difficult to solve this equation, 
\begin{eqnarray}\label{L25}
\eta(z,M)&=&-\frac{2}{\beta}\int_{z_{\omega}}^{z}dx\Bigl(2\omega e^x 
(\frac{\beta}{2}\cos\frac{\beta}{2}(x-z_{\omega}) 
-\frac{M}{2}\sin\frac{\beta}{2}(x-z_{\omega})) 
-\frac{E^{[1]}}{4}\sin\frac{\beta}{2}(x-z_{\omega})\Bigr) \nonumber \\ 
&&\times\Bigl(\sin\frac{\beta}{2}(x-z_{\omega})
\cos\frac{\beta}{2}(z-z_{\omega})-\cos\frac{\beta}{2}(x-z_{\omega})
\sin\frac{\beta}{2}(z-z_{\omega})\Bigr) \nonumber\\
&=& 
-\frac{1}{\beta}(\omega\beta(e^z(-\frac{\beta}{1+\beta^2}
\cos\frac{\beta}{2}(z-z_{\omega})
+\frac{1}{1+\beta^2}\sin\frac{\beta}{2}(z-z_{\omega})) \nonumber \\
&& +e^{z_{\omega}}(\frac{\beta}{1+\beta^2}\cos\frac{\beta}{2}(z-z_{\omega})
+\frac{1}{1+\beta^2}\sin\frac{\beta}{2}(z-z_{\omega}))
-(e^z-e^{z_{\omega}})\sin\frac{\beta}{2}(z-z_{\omega})) \nonumber \\
&& -\omega M (-e^z(\frac{1}{1+\beta^2}\cos\frac{\beta}{2}(z-z_{\omega})
+\frac{\beta}{1+\beta^2}\sin\frac{\beta}{2}(z-z_{\omega})) \nonumber \\
&& +e^{z_{\omega}}(\frac{1}{1+\beta^2}\cos\frac{\beta}{2}(z-z_{\omega})
-\frac{\beta}{1+\beta^2}\sin\frac{\beta}{2}(z-z_{\omega}))
+(e^z-e^{z_{\omega}})\cos\frac{\beta}{2}(z-z_{\omega})) \nonumber \\
&& -\frac{E^{[1]}}{4}((z-z_{\omega})\cos\frac{\beta}{2}(z-z_{\omega})
-\frac{1}{\beta}\sin\frac{\beta}{2}(z-z_{\omega})),
\end{eqnarray}
apart from the general solution, $\sin\frac{\beta}{2}(z-z_{\omega})$,
of the homogeneous equation. 

On the other hand, the descrete equation (\ref{L22}) with
$\omega{\cal J}^z=0$ is 
\begin{equation}\label{L26}
(4M{\cal J}^x-4({\cal J}^x)^2-E^{[0]})\sum_n \chi^{[1]}_{n,0}|n\rangle_{1}
=E^{[1]}\sum_n \chi^{[0]}_{n,0}|n\rangle_{1}.
\end{equation}
In order to solve Eq.(\ref{L26}), we use Eqs.(\ref{L9}) and (\ref{L16-1}) 
as in Eq.(\ref{L17}), and we have the following equation, 
\begin{equation}\label{L27}
(4MD_{w}-4D_{w}^{2}-E^{[0]})\bar{\chi}(w)=E^{[1]}\chi(w),
\end{equation} 
with 
$\bar{\chi}(w)=\sum_n \chi^{[1]}_{n,0}w^n$ and
$\chi(w)=\sum_n \chi^{[0]}_{n,0}w^n$.
Decomposing each power of $w$ in $\bar{\chi}(w)$, solution of 
Eq.(\ref{L27}) gives us solution of Eq.(\ref{L26}), $\chi^{[1]}_{n,0}$'s. 
Equation(\ref{L27}) can be solved exactly. 
The solution is given by 
\begin{equation}\label{L28}
\bar{\chi}(w)=E^{[1]}\int^{w}dv\; \frac{\hat{\psi}(v,\alpha_{+})
\hat{\psi}(w,\alpha_{-})
-\hat{\psi}(w,\alpha_{+})\hat{\psi}(v,\alpha_{-})}{\hat{\psi}(v,\alpha_{+})
\hat{\psi}'(v,\alpha_{-})
-\hat{\psi}'(v,\alpha_{+})\hat{\psi}(v,\alpha_{-})}
\Big(-\frac{\chi(v)}{(1-v^2)^2}\Big).
\end{equation} 
This can be verified by directly substituting (\ref{L28}) into Eq.(\ref{L27}).
Performing the integral in (\ref{L28}) and using the identity, 
\begin{equation}\label{L29}
\hat{\psi}(v,\alpha_{+})\hat{\psi}'(v,\alpha_{-})
-\hat{\psi}'(v,\alpha_{+})\hat{\psi}(v,\alpha_{-})
=-2i\beta\frac{1}{1-v^2}\hat{\psi}(v,\alpha_{+})\hat{\psi}(v,\alpha_{-}),
\end{equation}
we have
\begin{eqnarray}\label{L30}
\bar{\chi}(w)&=&
i\frac{E^{[1]}}{2\beta}
\int^{w}dv
\biggl(\bigl(\frac{\hat{\psi}(w,\alpha_{+})}{\hat{\psi}(v,\alpha_{+})}
-\frac{\hat{\psi}(w,\alpha_{-})}{\hat{\psi}(v,\alpha_{-})}\bigr)
\frac{\chi(v)}{1-v^2}
\biggr)\nonumber\\
&=&c_{+}\int^{w}dv\biggl(\bigl(\hat{\psi}(w,\alpha_{+})-
\frac{\hat{\psi}(v,\alpha_{+})}{\hat{\psi}(v,\alpha_{-})}
\hat{\psi}(w,\alpha_{-})
\bigr)\frac{1}{1-v^2}\biggr)
-(+ \leftrightarrow -),
\end{eqnarray}
where some factor in Eq.(\ref{L30}) can be simplified by
\begin{equation}\label{L31}
\frac{\hat{\psi}(v,\alpha_{+})}{\hat{\psi}(v,\alpha_{-})}
=\left(\frac{1-v}{1+v}\right)^{i\beta}.
\end{equation}
By using the identity,
\begin{eqnarray}\label{L32}
\int^{w}dv\biggl(\frac{1-v}{1+v}\biggr)^{i\beta}\frac{1}{1-v^2}
&=&-\frac{1}{2i\beta}\biggl(\frac{1-w}{1+w}\biggr)^{i\beta}+
\mbox{const.}\nonumber\\
\int^{w}dv\frac{1}{1-v^2}&=&\frac{1}{2}\ln|\frac{w+1}{w-1}|+\mbox{const.},
\end{eqnarray}
we can calculate the above integral exactly and we have
\begin{equation}\label{L33}
\bar{\chi}(w)=
\frac{E^{[1]}}{4i\beta}\ln|\frac{w-1}{w+1}|
(c_{+}\hat{\psi}(w,\alpha_{+})-c_{-}\hat{\psi}(w,\alpha_{-})),
\end{equation} 
up to the general solution $\chi(w)$. 
We hence obtain the exact form of $\chi^{[1]}_{n,0}$ as 
\begin{eqnarray}\label{L34}
\chi^{[1]}_{n,0}&=&\oint_{C}\frac{dw}{2\pi i}\frac{\bar{\chi}(w)}{w^{n+1}}
\nonumber\\
&=&\frac{E^{[1]}}{4i\beta}\sum_{l=0}^{n-1}\frac{1}{n-l}((-1)^{n-l}-1)
(c_{+}\psi_{l}(\alpha_+)-c_{-}\psi_{l}(\alpha_-)).
\end{eqnarray}

In order to compare the above result with the solution obtained by
the hard-wall approximation,
we approximate $\chi^{[1]}_{n,0}$ as follows,
\begin{eqnarray}\label{L35}
\chi^{[1]}_{n,0}
&\simeq& -\frac{E^{[1]}}{4i\beta}\sum_{l=0}^{n-1}\frac{1}{n-l}
\Bigl((c_{+}\frac{1}{lB(l,\frac{1}{2}-\alpha_{+})}-
       c_{-}\frac{1}{lB(l,\frac{1}{2}-\alpha_{-})}) \nonumber\\
&&\; \; -(-1)^{n}(c_{+}\frac{1}{lB(l,\frac{1}{2}+\alpha_{+})}-
          c_{-}\frac{1}{lB(l,\frac{1}{2}+\alpha_{-})})
\Bigr),
\end{eqnarray}
where we have picked out the leading terms in $\psi_l(\alpha_{\pm})$, i.e.,
\begin{equation}\label{L36}
\psi_l(\alpha)\simeq 
\frac{1}{lB(l,\frac{1}{2}-\alpha)}+(-1)^l\frac{1}{lB(l,\frac{1}{2}+\alpha)}.
\end{equation}
From the fact that $\frac{1}{n-l}\frac{1}{lB(l,\frac{1}{2}\pm\alpha)}$
has its maximum at $l=n-1$ and 
$\frac{1}{lB(l,\frac{1}{2}\pm\alpha)}$ is a slowly varying 
function for large $l$,
we find
\begin{eqnarray}\label{L37}
\sum_{l=0}^{n-1}\frac{1}{n-l}\frac{1}{lB(l,\frac{1}{2}\pm\alpha)}
&\simeq& \sum_{l=0}^{n-1}\frac{1}{n-l}\frac{1}{(n-1)B(n-1,
\frac{1}{2}\pm\alpha)}
\nonumber \\
&\simeq& (\ln n+\gamma)
\frac{\Gamma(n-\frac{1}{2}\pm\alpha)}{\Gamma(n)\Gamma
(\frac{1}{2}\pm\alpha)}\nonumber \\
&\simeq& (\ln n+\gamma)\frac{1}{\Gamma(\frac{1}{2}\pm\alpha)}
n^{-\frac{1}{2}\pm\alpha}
\end{eqnarray}
for large $n$, where $\gamma$ is Euler's gamma constant.
We therefore have the first order correction of $\chi_{n,0}$ as
\begin{eqnarray}\label{L38}
\chi^{[1]}_{n,0}&\simeq& -\frac{E^{[1]}}{4i\beta}(\ln n+\gamma)
\Bigl(n^{-\frac{1+M}{2}}(c_{+}n^{-\frac{\beta}{2}}-c_{-}n^{\frac{\beta}{2}})
+(-1)^n n^{-\frac{1-M}{2}}(c_{-}n^{-\frac{\beta}{2}}-c_{+}n^{\frac{\beta}{2}})
\Bigr)\nonumber \\
&=& -\frac{E^{[1]}}{4i\beta}(z+\gamma)\Bigl(
e^{-\frac{1+M}{2}z}\cos\frac{\beta}{2}(z-z_\omega)
+(-1)^{n+k+1}e^{-\frac{1-M}{2}z}\cos\frac{\beta}{2}(z-z_\omega)
\Bigr). \nonumber
\end{eqnarray}

In the limit $0 \ll z \ll z_{\omega}$ or $\omega e^z \ll 1$,
the solution in the hard-wall approximation is reduced to the form
\begin{equation}\label{L39}
e^{-\frac{1+M}{2}z}\bigl(\frac{E^{[1]}}{4\beta}z-\frac{\beta}{1+\beta^2}(1+M)-
\frac{E^{[1]}}{4\beta}z_{\omega}\bigr)\cos\frac{\beta}{2}(z-z_\omega)
+(-1)^n(M\leftrightarrow -M),
\end{equation}
which should be compared with the solution obtained from 
the difference equation
\begin{equation}\label{L40}
e^{-\frac{1+M}{2}z}\frac{E^{[1]}}{4\beta}(z+\gamma)
\cos\frac{\beta}{2}(z-z_\omega)
+(-1)^n(M\leftrightarrow -M).
\end{equation}
We hence obtain the `energy eigenvalue' in $O(\lambda)$,
\begin{equation}\label{L41}
E^{[1]}=-\frac{4\beta}{z_{\omega}+\gamma}\cdot\frac{\beta}{1+\beta^2}
(1+M).
\end{equation}
In the limit $M \rightarrow 0$ and $\omega \rightarrow 0$, we can 
neglect the higher-order terms in $\beta$ and we have 
\begin{equation}\label{L42}
E^{[1]}\simeq -\frac{4\beta^2}{z_{\omega}}= -\frac{4}{z_{\omega}^3}k^2.
\end{equation}
Thus the energy eigenvalue is 
\begin{eqnarray}\label{L43}
E_k&=&E_k^{[0]}+\lambda E_k^{[1]} \nonumber \\
   &=&\Big(\frac{\pi^2}{z_{\omega}^2}-\lambda
   \frac{4\pi^2}{z_{\omega}^3}\Big)k^2.
\end{eqnarray}

Since the one-particle Green's function has the form 
$\sum_{k}f(k)\exp(-E_k x)$ with
certain calculable functions $f(k)$,
the `imaginary-time' localization length $\xi_{\omega}$ is 
the inverse of the 
coefficient of $k^2$ in $E_k$, namely,
\begin{equation}\label{L44}
\xi_{\omega}=\frac{z_{\omega}^2}{\pi^2}\frac{1}{1-\frac{4\lambda}{z_{\omega}}}
\simeq \frac{z_{\omega}^2}{\pi^2}(1+\frac{4\lambda}{z_{\omega}}).
\end{equation}
The genuine localization length $\xi_{\epsilon}$ is the real part of 
$\xi_{\omega}$ after performing analytic continuation 
$i\omega\rightarrow\epsilon+i\omega$, and then Eq.(\ref{L44}) leads to
\begin{equation}\label{L45}
\xi_{\epsilon}\simeq \frac{1}{g}\Big(\frac{|\ln\frac{\epsilon}{2g}|^2}{\pi^2}+
g\lambda\frac{4|\ln\frac{\epsilon}{2g}|}{\pi^2}\Big),
\end{equation}
where we have recovered the constant $g$ and use the fact that 
$\ln\frac{\epsilon}{g}\simeq\ln\frac{\epsilon}{2g}$ for small $\epsilon$.
Equation(\ref{L45}) shows that the localization length $\xi_{\epsilon}$
is a increasing function of $(g\lambda)$.
This means that the nonlocally-correlated disorder enhances 
the delocalization.
Moreover the number of states with energy below $\epsilon$ is 
\begin{eqnarray}\label{L46}
N(\epsilon)&=&\int_0^{\epsilon}d\epsilon\mbox{ }\rho(\epsilon) \nonumber \\
           &=& \frac{g}{2}\Bigl(\frac{1}{|\ln\frac{\epsilon}{2g}|^2}-
               \frac{4g\lambda}{|\ln\frac{\epsilon}{2g}|^3}\Bigr)
               +\mbox{(higher-order terms)},
\end{eqnarray}
and this shows that the product of the localization length 
and the number of states is constant up to the first order
of $(g\lambda)$, 
\begin{equation}\label{L47}
\xi_{\epsilon}N(\epsilon)=\frac{1}{2\pi^2}+O((g\lambda)^2).
\end{equation}
Physical meaning of this result will be discussed in Sect.4
from microscopic point of view (see also Ref.\cite{TTIK}).

\setcounter{equation}{0}
\section{Discussion}
In this paper, we have studied how the nonlocally 
correlated disorder affects the the mean
localization length $\xi_{\epsilon}$ in the random-mass Dirac fermions
by making use of SUSY methods.
(In fact this is a low-energy effective model of the random hopping tight
binding model\cite{BF}.)
We found that $\xi_{\epsilon}$ is an increasing function of
the correlation length $\lambda$. 
This result is consistent with the behavior of 
the DOS $\rho(\epsilon)$ obtained
in the previous paper\cite{IK} and the numerical studies in Ref.\cite{TTIK}.
There exists a simple intuitive picture of the result: 
For the case of larger correlation length $\lambda$, average distance
between kinks is longer and also average height of kinks
is lower.
Therefore, finite correlation of disorders hinders the scattering
of fermions by disorders and enhances the delocalization of fermions.
Though this result is obtained in the specific one-dimensional model,
we expect that similar behaviours hold in general 
random-disordered systems.

Let us consider relationship between the localization
length $\xi_{\epsilon}$ and the number of states $N(\epsilon)$.
Our SUSY calculation suggests that the following relation
\begin{equation}
\xi_{\epsilon}=\frac{\mbox{const.}}{N(\epsilon)}
\end{equation}
holds (up to O($g\lambda$)). This can be understood as below.
As is shown by the numerical calculations of wave
functions of random-mass Dirac fermions\cite{TTIK}, 
fermions in the random potential
localize within an interval between adjacent nodes.
According to `one-dimensional node counting theorem'\cite{CDM}, the
number of nodes
$N_{d}(\epsilon)$ is equal to $LN(\epsilon)-1$.
Let us suppose that the position of each node $x_{i} (i=1,2,\cdots,N_{d})$
is randomly distributed under the constraint
$0<x_{1}<\cdots <x_{i}<x_{i+1}<\cdots <x_{N_{d}}<L$.
The expectation value $\langle x_{i}\rangle$ is then given as
\begin{eqnarray}
\langle x_{i}\rangle&=&\frac{1}{{\cal
N}}\prod_{k=1}^{N_{d}}\int_{0}^{L}\frac{dx_{k}}{L}
x_{i}\prod_{l=1}^{N_{d}-1}\theta(x_{l+1}-x_{l})\nonumber \\
&=&L\frac{i}{N_{d}+1}=\frac{i}{N(\epsilon)},
\end{eqnarray}
with the normalization constant
\begin{equation}
{\cal N}=\prod_{k=1}^{N_{d}}\int_{0}^{L}\frac{dx_{k}}{L}
\prod_{l=1}^{N_{d}-1}\theta(x_{l+1}-x_{l}).
\end{equation}
Therefore it is found that the expectation value of the distance between
adjacent nodes is
$\langle x_{i+1}\rangle-\langle x_{i}\rangle=\frac{1}{N}$.
According to this estimation, we roughly understand that
the localization length is to be inversely proportional
to the number of states $N(\epsilon)$.

In usual cases, it is not easy to obtain the localization length 
accurately by numerical calculations.
However very recently, a very useful method for that
was proposed, i.e., the non-Hermitian extension by introduction of
an imaginary vector potential\cite{HN}.
Then we hope that the localization length in the present model
is also calculated numerically by that method, and the analytical expression,
which we obtained in this paper, is compared with numerical calculations.
This is under study and results will be reported in a future publication.

While the results in this paper are obtained
 for the (effective) random hopping 
tight binding (RHTB) model, they give
some important implications for closely related model --- spin-Peierls model
\cite{IAF}. Since the $z$-component of spin in the spin-Peierls
model corresponds to the presence or absence of the tight-binding fermion,
the Green's function in the RHTB model can be considered as 
a correlation function of spins in the spin-Peierls model.
In fact, the single-fermion Green's function of the RHTB model,
\[\int_{0}^{\infty}dt\;  e^{i\epsilon t}
\langle v| c_n(t)c^{\dagger}_{n'}(0)|v\rangle,\]
coincides
with the spin-spin correlation function
\[\int_{0}^{\infty}dt \; e^{i\epsilon t}
\langle v| S^{-}(n,t)S^{+}(n',0)|v\rangle, \]
where $S^{\pm}(n,0)$ are the spin-up and spin-down operators at site
$n$, and $S^{-}(n,t)$ is given by
\[S^{-}(n,t)=e^{i{\cal H}_{SP}t}S^{-}(n,0)e^{-i{\cal H}_{SP}t}\]
with the spin-Peierles Hamiltonian ${\cal H}_{SP}$.
The operators $S^{+}(n,0)$ and $c_{n}$ are related with each other by the
Jordan-Wigner transformation.
In the spin systems, the mean localization length $\xi_{\epsilon}$,
which is calculated in this paper,
can be considered as the mean correlation length of the spins.
We therefore find that the spin-spin correlation is enhanced by 
the suppression of randomness
of disorders, because the suppression enhances the (quasi-)extended
states near the band center $\epsilon =0$. 

Let us compare the above result with the experimental
observation in Ref.\cite{HA2}. 
In Ref.\cite{HA2}, a phase diagram of
$\mbox{Cu}_{1-x}\mbox{Zn}_{x}\mbox{GeO}_{3}$ was
obtained by the measurements of the magnetic susceptibility. 
The phase diagram shows that the antiferromagnetic
(AF) order is enhanced by decrease of the impurity concentration at
higher-doped region ($x>0.03$). (Note that the spin-Peierls state collapses
at around $x=0.03$\cite{HA2}.)
We expect that the white-noise limit $\lambda=0$
describes the system at around
$x \sim 0.5$, i.e., extremely high-doped region.
As increasing $\lambda$, we approach the lower-doped region.
Then as the calculations in this paper show, the extended low-energy
excitations are enhanced as $\lambda \rightarrow$ large.
This fact obviously means the enhancement of the AF order 
in the lower-doped region.
This result is 
qualitatively in good agreement with the experiments.
It is interesting to give more quantitative argument about this problem.
This can be a useful check on our results.


\end{document}